\documentclass[12pt]{book}

\usepackage[dvips]{graphicx,color}
\usepackage{makeidx,high}

\def\la{\hbox{{\lower -2.5pt\hbox{$<$}}\hskip -8pt\raise
-2.5pt\hbox{$\sim$}}}
\def\ga{\hbox{{\lower -2.5pt\hbox{$>$}}\hskip -8pt\raise
-2.5pt\hbox{$\sim$}}}
\def\ltsima{$\; \buildrel < \over \sim \;$}
\def\simlt{\lower.5ex\hbox{\ltsima}}
\def\gtsima{$\; \buildrel > \over \sim \;$}
\def\simgt{\lower.5ex\hbox{\gtsima}}

\makeauthorindex
\makeindex

\newcommand{\dsc}[4]{$#1\pm#2\,(#3)\,$\boldmath{$[#4]$}}

\begin{document}

\BookTitle{\itshape Extremely High Energy Cosmic Rays}
\CopyRight{\copyright 2002 by Universal Academy Press, Inc.}
\pagenumbering{arabic}

\chapter{Low Statistics of EHECRs}

\author{%
Angela V. OLINTO\\
{\it Department of Astronomy \& Astrophysics,  \& Enrico Fermi
Institute, \\ The University of Chicago, Chicago, IL 60637, USA}\\
olinto@oddjob.uchicago.edu\\
Daniel DE MARCO\\
INFN \& Universit\`a degli Studi di Genova\\ Via
Dodecaneso, 33 -
16100 Genova, ITALY\\
ddm@ge.infn.it\\
Pasquale BLASI\\
INAF/Osservatorio Astrofisico di Arcetri\\ Largo E. Fermi, 5 -
50125 Firenze, ITALY\\
blasi@arcetri.astro.it}

\AuthorContents{A. V. \ Olinto} 
\AuthorIndex{Olinto}{A. V.} 

\section*{Abstract}
The nature of the unknown sources of ultra-high energy cosmic rays can be
revealed through the detection of the GZK feature in the cosmic ray
spectrum. The only two experiments that have probed this energy range, AGASA
and HiRes, have apparently conflicting results. HiRes measured a flux
consistent with the GZK feature while AGASA reported a larger than expected
flux of so-called Super-GZK particles. Here we emphasize that neither
experiment has gathered the statistics necessary for making a definitive
measurement of the GZK cutoff. The photo-pion production responsible for
the GZK feature is stochastic for energies around the cutoff leading to
large fluctuations of the spectrum for low statistics measurements. We
show that the results from AGASA and HiRes results are within about $ 2
\sigma$ of one another by simulating 400 spectra for a range of input
spectral indices normalized to the number of events above $10^{19}$ eV for
each experiment. If a 15\% systematic correction in energy is applied to
both experiments, the agreement between the experiments improves
considerably and the best fit input spectral index becomes $\sim 2.6$ for
both data sets. Our results clearly show
the need for much larger experiments such as Auger and EUSO, that
can increase the number of detected events by 2 orders of magnitude. Only
large statistics experiments can finally prove or disprove the  existence
of the GZK feature in the cosmic ray spectrum.

\section{Introduction}

The most extreme accelerators in the universe remain a mystery. Particles
with energies above $10^{20}$ eV have been observed to arrive on Earth from
directions in space that are not clearly associated with any particular
source. The mystery of the origin and nature of these particles should be
resolved once we have a statistically significant measurement of the
spectrum and angular distribution of these particles. Next generation
experiments such as the Pierre Auger Project and the EUSO experiment should
be able to finally resolve this mystery. 

Presently available data show a number of interesting hints of possible
resolutions to this mystery. The High Resolution Fly's Eye
fluorescence experiment has reported hints of a flux decrease around the
Greisen-Zatsepin-Kuzmin (GZK) feature. A significant decrease in the flux
of ultra-high energy extragalactic protons was predicted by Greisen
\cite{greisen} and Zatsepin and Kuzmin \cite{zatkuz} following the discovery
of the cosmic microwave background (CMB). Protons with energies above $7
\times 10^{19}$ eV lose a significant fraction of their energy through the
photo-pion production off the CMB as they traverse tens of Mpc in the
universe. If the HiRes result is confirmed with much better statistics the
nature of the highest energy cosmic rays are most likely to be
extragalactic protons. 

The giant airshower ground array, AGASA, does not see the suppression
expected at the GZK feature. They reported about 11 events above $10^{20}$
eV implying a flux above the expected level due to photo-pion
production. The excess flux seems incompatible with extragalactic protons
and a number of alternatives have been proposed to address this discrepancy
\cite{reviews}. In addition, the distribution in the sky of arrival
directions reported by AGASA shows hints of clustering in small scales.
These may be the first hints of point sources. 

Here I discuss the significance of the apparent inconsistencies between the
two experiments and re-emphasize the lack of sufficient statistics in
both experiments to address the very important question of the origin of
the highest energy particles ever observed. This presentation is based
on results reported in the recent work of DeMarco, Blasi, and
Olinto (2003)\cite{dbo03} (see also \cite{bbo00}).

\section{AGASA versus HiResI}

\begin{figure}
\begin{center}
\includegraphics[width=0.9\textwidth]{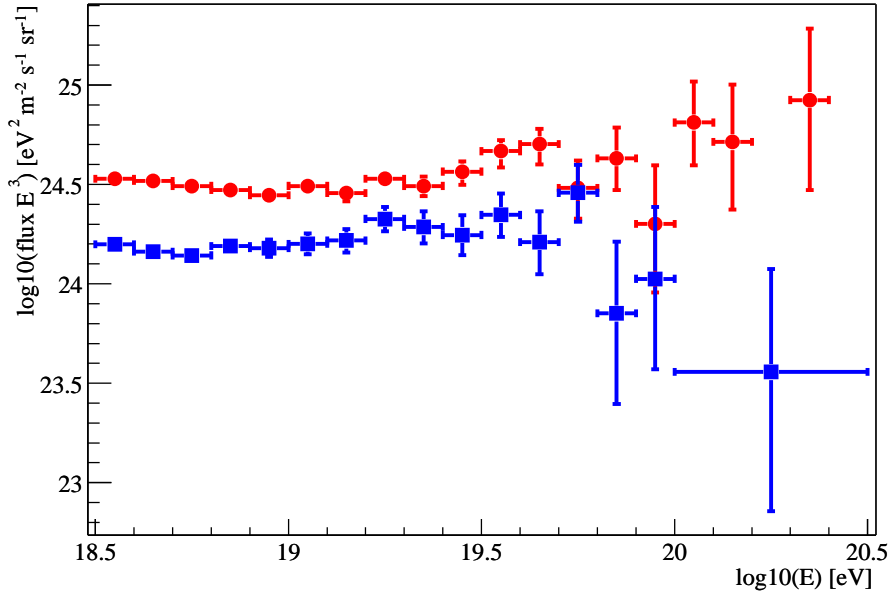}
\caption{Flux of cosmic rays multiplied by $E^3$ versus the energy,
$E$. Circles show the AGASA spectrum from
\cite{AGASA} while squares show the  HiResI spectrum from \cite{HIRES2}.}
\label{fig:agasahires}
\end{center}
\end{figure}

The two largest experiments to measure the flux of UHECRs, AGASA
\cite{AGASA} and HiResI \cite{HIRES2},  reported apparently conflicting
results shown in Fig.
\ref{fig:agasahires}
The figure shows that the HiResI data are systematically below
AGASA data and that HiResI sees a suppression at $\sim 10^{20}$ eV
that resembles the GZK feature while AGASA does not.
The number of events above $10^{19}$ eV, $10^{19.6}$ eV and
$10^{20}$ eV for AGASA and HiResI are shown in Table
\ref{tab:data_shifted}
We also show in the same
table the number of events above the same energy thresholds,
when a systematic energy decrease of 15\% is applied to the AGASA data
while a systematic energy increase of $15\%$ is applied to the HiResI
data. These systematic energy shifts are within the reported
uncertainties in energy measurements in both experiments. (See this
proceedings for a discussion of energy errors in AGASA \cite{nagano} and
HiRes \cite{marteens}.)
\begin{table}
\begin{center}
\begin{tabular}{c|cc|cc}
$\log(E_{th})\,({\rm eV})$ &
\multicolumn{2}{c}{AGASA} &
\multicolumn{2}{|c}{HiResI} \\
& data & -15\% & data & +15\% \\
\hline
19  & 866 & 651 & 300 & 367\\
19.6 & 72 & 48 & 27 & 39\\
20 & 11 & 7 & 1 & 2.2\\
\hline
\end{tabular}
\caption{Number of events for AGASA and HiResI detected above the
energy thresholds
reported in the first column.}
\label{tab:data_shifted}
\end{center}
\end{table}

In order to understand the difference between AGASA and HiRes
data, we first determine the best fit injection spectrum for each data set
separately by running 400 simulations of extragalactic proton propagation
normalizing to the  {\it data} column in Table
\ref{tab:data_shifted} The injection spectrum is taken to be a power law
with index $\gamma$ between 2.3 and 2.9 with steps of 0.1. For each
injection spectrum we calculated the $\chi^2$ indicator (averaged over 400
realizations for each injection spectrum). The errors used for the
evaluation of the $\chi^2$ are due  to the square roots of the number of
observed events. As reported in Table \ref{tab:chiq}, the best fit injection
spectrum can change appreciably depending on the minimum energy above which
the fit is calculated. In these tables we
give  $\chi_{e}^2(N)$, where $N$ is the number of degrees of  freedom,
while the subscript, $e$, is the logarithm of $E_{th}$ (in eV), the energy
above which the fit has been calculated. The numbers in bold-face correspond
to the best fit injection spectrum.  These fits are dominated by the low
energy data rather than by the poorer statistics at the higher energies.

\begin{table}
\begin{center}
\begin{tabular}{l|ccc|ccc}
&\multicolumn{3}{c|}{AGASA}&\multicolumn{3}{c}{HiResI}\\
$\gamma$    &   $\chi^2_{18.5}(17)$ &   $\chi^2_{19}(12)$   &
$\chi^2_{19.6}(6)$  &   $\chi^2_{18.5}(15)$ &   $\chi^2_{19}(10)$   &
$\chi^2_{19.6}(4)$\\
\hline
2.3 & 1694 & 34       & 17          & 145     & 29      & 23\\
2.4 & 1215 & 24       & 12          & 80      & 19      & 15\\
2.5 & 724  & 19       & {\bf 10}    & 36      & 14      & 11\\
2.6 & 248  & {\bf 16} & {\bf 10}    & {\bf 14}& 9       & 7\\
2.7 & 82   & 17       & 11          & 33      & {\bf 7} & 5\\
2.8 & {\bf 80}& 21    & 13          & 126     & {\bf 7} & {\bf 4}\\
2.9 & 316  & 27       & 15          & 257     & 9       & {\bf 4}\\
\hline
\end{tabular}
\caption{$\chi^2$ for fits to AGASA and HiResI data above $10^{18.5}$ eV,
$10^{19}$ eV, and $10^{19.6}$ eV for varying spectral index $\gamma$. In
parenthesis the number of degrees of freedom.}\label{tab:chiq}
\end{center}
\end{table}

For energies above $10^{18.5}$ eV, the best fit spectra are $E^{-2.8}$ for
AGASA and
$E^{-2.6}$ for HiRes. If the data at energies above $10^{19}$ eV are
used for the fit, the best fit injection spectrum is $E^{-2.6}$ for
AGASA and between $E^{-2.7}$ and $E^{-2.8}$ for HiRes. If the fit is
carried out on the highest energy data ($E>10^{19.6}$ eV), AGASA
requires an injection spectrum between $E^{-2.5}$ and $E^{-2.6}$, while
$E^{-2.8}$ or $E^{-2.9}$ fits the HiRes data in the same energy region.
Note that the two sets of data uncorrected for any possible systematic
errors require different injection spectra that change with  $E_{th}$.

\begin{table}
\begin{center}
\begin{tabular}{c|c|c|c}
${E_{th} \over {\rm eV}}$& $\gamma=2.5$ & $\gamma=2.6$ & $\gamma=2.8$\\
\hline
$10^{19.6}$ & \dsc{65}{8.2}{+0.5}{+0.3} &  \dsc{58}{7.6}{+1.4}{+1.0} & 
\dsc{46}{6.8}{+2.8}{+2.2} \\ $10^{20}$   & \dsc{3.5}{1.9}{+2.4}{+2.1} &
\dsc{2.8}{1.7}{+2.6}{+2.3} & \dsc{2.0}{1.4}{+2.8}{+2.6} \\
\hline
\end{tabular}
\caption{Number of events expected above $E_{th}$ for different
injection spectra assuming the AGASA statistics above $10^{19}$ eV.
In parenthesis are the number of standard deviations, $\sigma$, between the 
expected number of events and the observed number.  In square brackets are
the combined error bar between simulations and observations,
$\sigma_{tot}$.}
\label{tab:discrepanza_agasa}
\end{center}
\end{table}

\begin{table}
\begin{center}
\begin{tabular}{c|c|c|c}
${E_{th}\over {eV}}$ & $\gamma=2.6$ & $\gamma=2.7$ & $\gamma=2.8$\\
\hline
$10^{19.6}$ & \dsc{31}{5.6}{-0.8}{-0.6} & \dsc{28}{5.3}{-0.2}{-0.1} &
\dsc{26}{5.2}{+0.3}{+0.2} \\ $10^{20}$ & \dsc{1.9}{1.4}{-0.9}{-0.5} &
\dsc{1.5}{1.2}{-0.5}{-0.3} & \dsc{1.3}{1.2}{-0.3}{-0.2} \\
\hline
\end{tabular}
\caption{Number of events expected above $E_{th}$ for different
injection spectra assuming the HiResI statistics above $10^{19}$ eV from Table
\ref{tab:data_shifted}. In parenthesis are the number of standard deviations,
$\sigma$, between the  expected number of events and the observed number.  In
square brackets are the combined error bar between simulations and
observations,
$\sigma_{tot}$.}
\label{tab:discrepanza_hires}
\end{center}
\end{table}

In order to quantify the significance of the detection or lack of
the GZK flux suppression, in Tables \ref{tab:discrepanza_agasa} and
\ref{tab:discrepanza_hires} we report the expected (namely simulated)
number of events above the energy threshold indicated for different
injection specta and the discrepancy between the simulations and the data in
Table \ref{tab:data_shifted} (in parenthesis) in numbers of standard
deviations, $\sigma$, after comparison with the data. From Tables
\ref{tab:discrepanza_agasa} and
\ref{tab:discrepanza_hires} we can see that while HiResI is consistent with
the existence of the GZK feature in the spectrum of UHECRs, AGASA
fails to detect the same feature, but only at the $\sim 2.5\sigma$ level for
the best fit injection spectra.

\begin{figure}
\begin{center}
\includegraphics[width=0.7\textwidth]{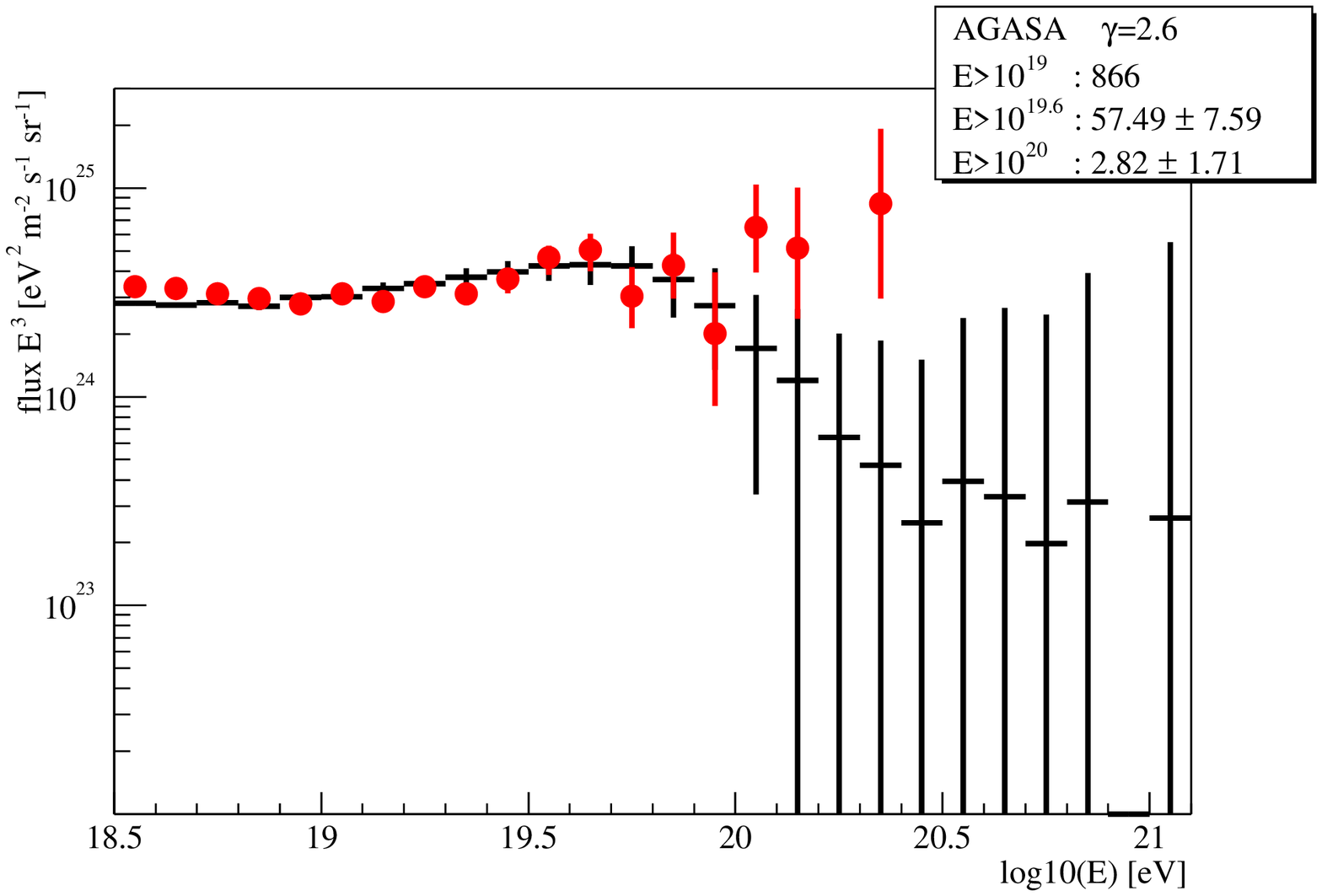}
\caption{Simulations of AGASA statistics with injection spectra
$E^{-2.6}$ (upper plot) and $E^{-2.8}$ (lower plot). The crosses with
error bars are the results of simulations, while the grey
points are the AGASA data.}\label{fig:agasa_2627}
\end{center}
\end{figure}

\begin{figure}
\begin{center}
\includegraphics[width=0.7\textwidth]{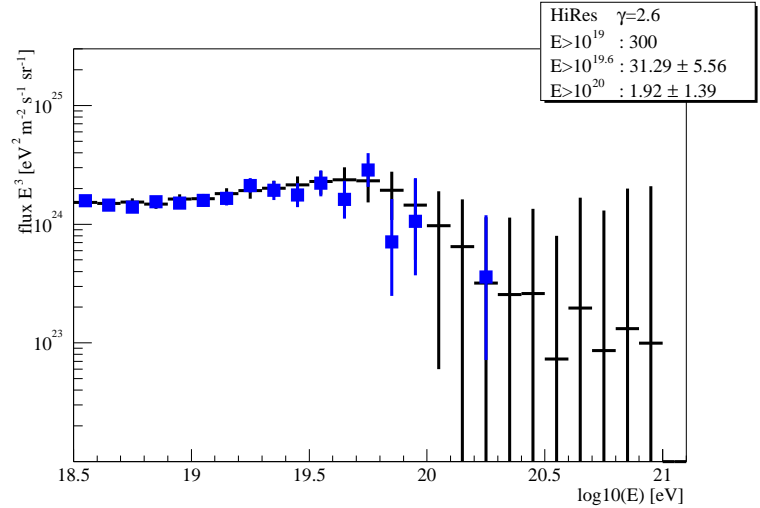}
\caption{Simulations of HiRes statistics with injection spectra
$E^{-2.6}$ (upper plot) and $E^{-2.7}$ (lower plot). The crosses with
error bars are the results of simulations, while the squares
are the HiRes data.}\label{fig:hires_2528}
\end{center}
\end{figure}

In Figs. \ref{fig:agasa_2627} and \ref{fig:hires_2528} we show the
uncertainties in the predicted fluxes for AGASA and HiRes respectively, for
a injection spectrum with spectral index 2.6. These plots show clearly the
low level of significance that the detections above $E_{GZK}$ have with low
statistics.  The large error bars that are generated by our simulations at
the high energy end of the spectrum are mainly due to
the stochastic nature of the process of photo-pion production: in some
realizations some energy bins are populated by a few events, while in
others the few particles in the same energy bin do not
produce a pion and get to the observer unaffected. The large
fluctuations are unavoidable with the extremely small statistics available
with present experiments. On the other hand, the error bars at lower energies
are minuscule, so that the two data sets (AGASA and HiResI) cannot be
considered to be two different realizations of the same phenomenon. The
discrepancies at the lower energy range must be due to  systematic errors in
at least one if not both experiments.

Taking into account the (theoretical) error bars determined by the
simulations, we report in Tables \ref{tab:discrepanza_agasa} and
\ref{tab:discrepanza_hires} the standard deviations due to the low
statistics ($\sigma_{sim}$). The observed number of events above the same
threshold also has a observational error bar
$\sigma_{obs}$ associated with that enregy range. The combined error 
can be estimated by $\sigma_{tot}=(\sigma_{sim}^2+
\sigma_{obs}^2)^{1/2}$. We show in Tables
\ref {tab:discrepanza_agasa} for AGASA and
\ref {tab:discrepanza_hires} for HiRes the standard deviations in the
simulations ($\sigma_{sim}$) in parenthesis  and the standard deviation with
the combined error in units of $\sigma_{tot}$ in square brackets. The
effective discrepancy between predictions and the AGASA data is at the
level of $2.1-2.5\sigma$. A definitive answer to the question of
whether the GZK feature is there or not awaits a significant improvement in
statistics at high energies.

As seen in Fig. \ref{fig:agasahires}, the difference between the AGASA
and HiResI spectra is not only in the presence or absence
of the GZK feature: the two spectra, when multiplied by $E^3$, are
systematically shifted by about a factor of two. This shift suggests that
there may be a systematic errors either in the energy or the flux determination
of at least one of the two experiments. Possible systematic effects are
discussed in \cite{nagano} and \cite{astro0209422} for the AGASA
collaboration and in
\cite{marteens} and \cite{HIRES2} for HiResI. At the lower end of the energy
range the offset may be due to the notoriously difficult determination of
the AGASA aperture at threshold while the discrepancies at the highest
energies is unclear at the moment. In any case, a systematic error of $\sim
15\%$ in the energy determination is well within the limits of systematic
errors reported by both collaborations.

In order to illustrate the difficulty in determining the
existence of the GZK feature in the observed data in the
presence of systematic errors, we split the energy gap by assuming that
the energies as determined by the AGASA collaboration
are overestimated by $15\%$, while the HiRes energies
are underestimated by the same amount. The number of events
above an energy threshold is again reported in Table
\ref{tab:data_shifted}. The values of $\chi^2$ obtained in this case are
reported in Table  \ref{tab:chiq_shift}.
\begin{table}
\begin{center}
\begin{tabular}{l|ccc|ccc}
&\multicolumn{3}{c|}{AGASA}&\multicolumn{3}{c}{HiResI}\\
$\gamma$    &   $\chi^2_{18.6}(15)$ &   $\chi^2_{19}(11)$   &
$\chi^2_{19.6}(5)$ &   $\chi^2_{18.6}(14)$ &   $\chi^2_{19}(10)$   &
$\chi^2_{19.6}(4)$\\
\hline
2.3 & 505     & 18       & 12       & 79       & 13    & 7\\
2.4 & 351     & 13       & 8.5      & 40       & 7     & 4\\
2.5 & 188     & {\bf 9}  & {\bf 5.6}& 13       & 3.7     & 2.0\\
2.6 & 54      & {\bf 9}  & {\bf 5.6}& {\bf 6}& {\bf 2.0} & {\bf 1.1}\\
2.7 & {\bf 20}& 11       & 6.4      & 23       & 3.1     & 1.4\\
2.8 & 54      & 15       & 7.2      & 94       & 6     & 2.4\\
2.9 & 145     & 20       & 9.1      & 176      & 10    & 4\\
\hline
\end{tabular}
\caption{$\chi^2$ for AGASA and HiResI in which a correction for a systematic
$15\%$ overestimate of the energies has been assumed for AGASA and a
$15\%$ underestimate of the energies has been assumed for HiResI.}
\label{tab:chiq_shift}
\end{center}
\end{table}

For AGASA, the best fit injection spectrum is now between $E^{-2.5}$
and $E^{-2.6}$ above $10^{19}$ eV and above $10^{19.6}$ eV (the $\chi^2$
values are identical). For the HiRes data,
the best fit injection spectrum is $E^{-2.6}$ for the whole set
of data, independent of the threshold. It is interesting to note that the
best fit injection spectrum appears much more stable after the
correction of the $15\%$ systematics has been carried out. Moreover,
the best fit injection spectra as derived for each experiment
independently coincides for the corrected data unlike the uncorrected case.
This suggests that combined systematic errors in the energy determination 
at the $\sim$ 30\% level may in fact be present.

The expected numbers of events with energy above $10^{19.6}$ eV
and above $10^{20}$ eV with the deviation from the data are
reported in Tables \ref{tab:discrepanza_agasa085} and
\ref{tab:discrepanza_hires115}: while HiResI data remain in
agreement with the prediction of a GZK feature, the AGASA
data seem to depart from such prediction but only at the
level of $\sim 1.8\sigma$. The systematics in the energy
determination further decreased  the significance
of the GZK feature, such that the AGASA and
HiResI data are in fact only less than $2\sigma$ away from each other.
\begin{table}
\begin{center}
\begin{tabular}{c|c|c|c}
${E \over {eV}}$ & $\gamma=2.5$ & $\gamma=2.6$ & $\gamma=2.7$ \\
\hline
$10^{19.6}$ & \dsc{49}{6.9}{+0.2}{+0.1} & \dsc{43}{6.5}{+0.8}{+0.5} &
\dsc{39}{6.1}{+1.3}{+1.0} \\ $10^{20}$   & \dsc{2.6}{1.6}{+1.7}{+1.4} &
\dsc{2.3}{1.5}{+1.8}{+1.5} & \dsc{1.8}{1.4}{+2.0}{+1.7} \\ 
\hline
\end{tabular}
\caption{Expected number of events above $E_{th}$ when a systematics of
$-15\%$ is introduced in the energy determination of AGASA. The number of
standard deviations between simulations and observations , $\sigma$, is listed
in parenthesis. In square brackets are the discrepancies calculated
with a combined error bar of simulation and observation uncertainties,
$\sigma_{tot}$.}\label{tab:discrepanza_agasa085}
\end{center}
\end{table}

\begin{table}
\begin{center}
\begin{tabular}{c|c|c}
${E \over {eV}}$&$\gamma=2.5$ & $\gamma=2.6$ \\
\hline
$10^{19.6}$ & \dsc{43}{6.3}{-0.6}{-0.4} & \dsc{38}{6.0}{+0.1}{+0.1} \\
$10^{20}$   & \dsc{2.8}{1.7}{-0.4}{-0.3} & \dsc{2.3}{1.5}{-0.1}{-0.1} \\
\hline
\end{tabular}
\caption{Expected number of events above $E_{th}$ when a systematics of
$+15\%$ is introduced in the energy determination of HiResI. The number of
standard deviations between simulations and observations , $\sigma$, is listed
in parenthesis. In square brackets are the discrepancies calculated
with a combined error bar of simulation and observation uncertainties,
$\sigma_{tot}$.}\label{tab:discrepanza_hires115}
\end{center}
\end{table}

We can use the same procedure illustrated above to
evaluate the effect of the error bars in the simulated data compared to the
data corrected by 15\%. The results are reported in square brackets in Tables
\ref {tab:discrepanza_agasa085} (for AGASA) and \ref {tab:discrepanza_hires115}
(for HiRes), showing that the effective discrepancy between expectations (with
uncertainties due to discrete energy losses and cosmic variance) and AGASA
data is even smaller, only at the $1.5\sigma$ level. 

\section{Conclusions}

The statistical significance of the UHECR spectra measured by the two
largest experiments in operation, AGASA and HiRes, is not sufficient to
determine the nature and origin of these particles. The
discrepancies between the two experiments is below 2.6 $\sigma$, but AGASA
sees hints of a flux above the predicted GZK suppression and hints of
point sources in arrival direction distribution while HiRes sees hints of
the GZK feature. If confirmed by future experiments with much larger
statistics, the increase in flux  relative the GZK prediction hinted by
AGASA would be of great interest. This may signal the presence of a new
component at the highest energies that compensates for the expected
suppression due to photo-pion production, or the effect of new physics in
particle interactions.

Identifying the cause of the systematic energy and/or flux shift between
the AGASA and the HiRes spectra is crucial for understanding the nature
of UHECRs. This discrepancy has stimulated a number of efforts  to search
for the source of these systematic errors including the construction of
hybrid detectors, such as Auger, that utilize both ground arrays and
fluorescence detectors. 
If systematic errors of about 15\% are present in both
experiments the two spectra can be shifted into agreement with a best fit
injection spectrum of $E^{-\gamma}$ with $\gamma=2.6$.

\begin{figure}
\begin{center}
\includegraphics[width=0.7\textwidth]{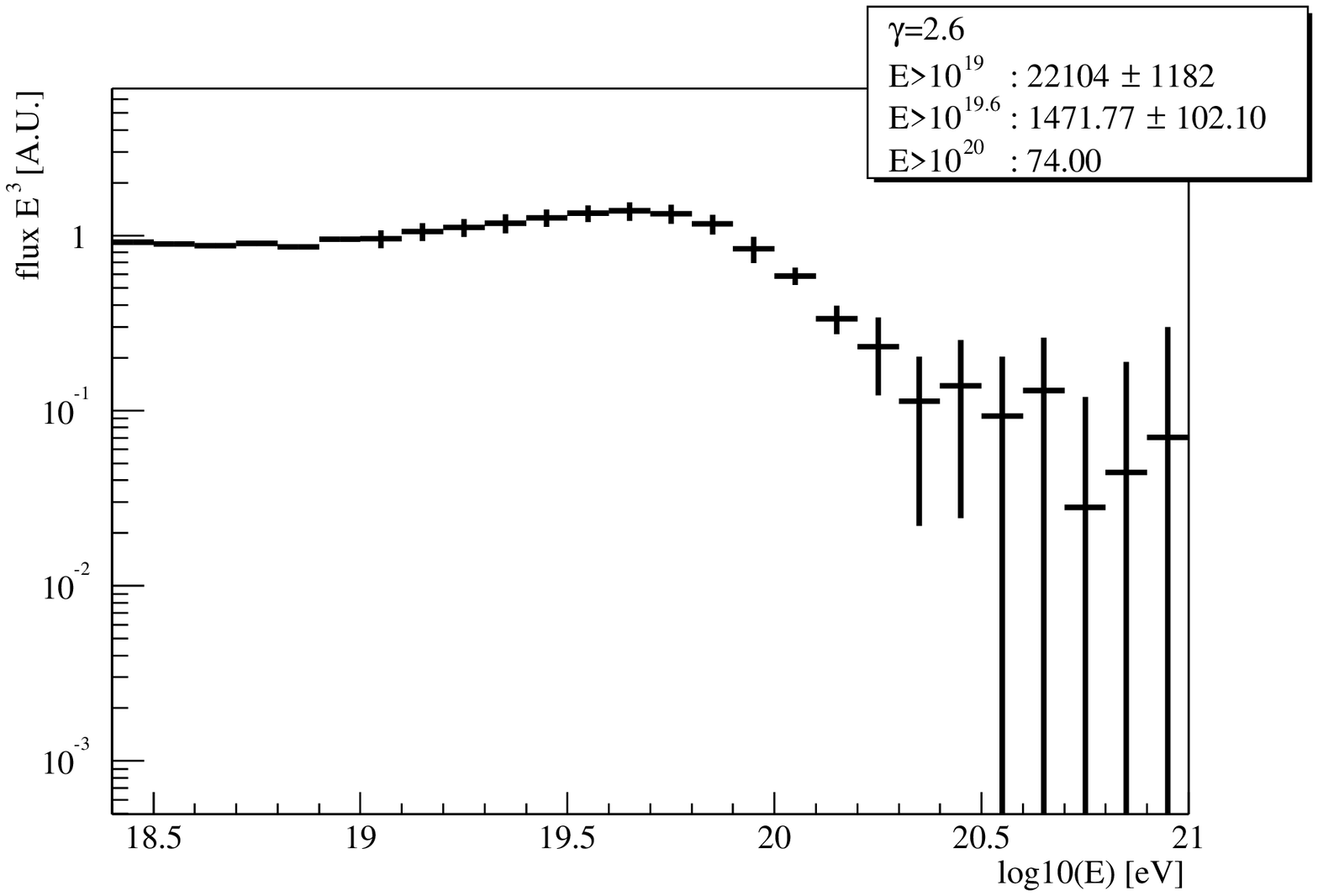}
\includegraphics[width=0.7\textwidth]{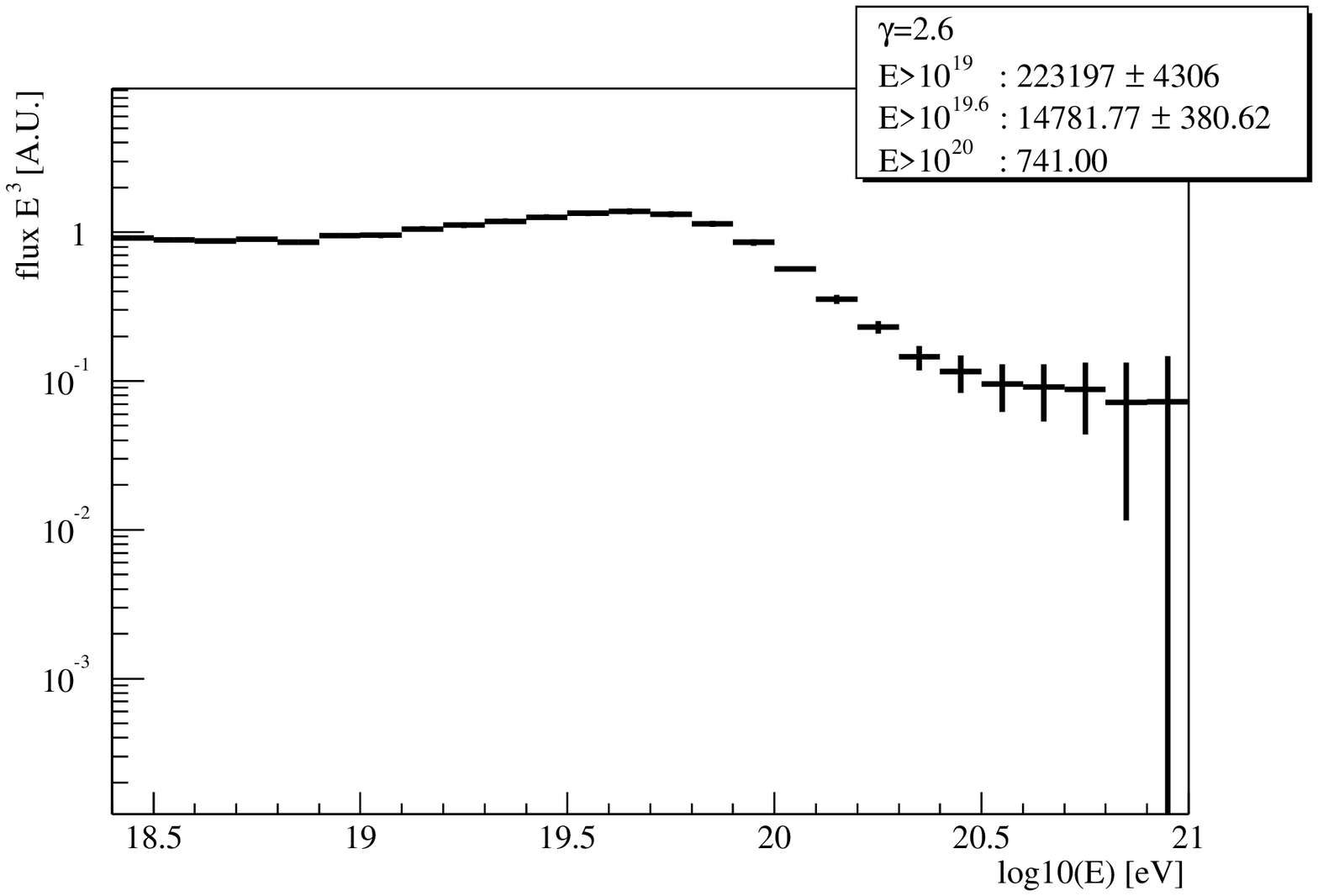}
\caption{Predicted spectra and error bars for 3 years of operation of
Auger (upper plot) and EUSO (lower plot).}
\label{fig:newgen}
\end{center}
\end{figure}

With the low statistical significance of the possible discrepancy
between AGASA and the expected fluxes (or the HiResI data), it is inaccurate
to claim either the detection of the GZK feature or the extension of
the UHECR spectrum beyond $E_{GZK}$ at this point in time. A new generation
of experiments is needed in order to finally give the proper answer
to this question.
In Fig. \ref{fig:newgen} we report the simulated spectra that should be
achieved in 3 years of operation of Auger (upper panel) and EUSO (lower
panel). The error bars reflect the fluctuations about the mean spectrum
with an injection spectrum $E^{-2.6}$. The energy region
where statistical fluctuations dominate the spectrum is moved to $\sim
10^{20.6}$ eV for Auger \cite{auger}, allowing a clear identification of the
GZK feature. This {\it fluctuations} dominated region stands beyond
$10^{21}$ eV for EUSO \cite{EUSO}. 

\subsection*{Acknowledgements}
 We thank the organizers of the International workshop on Extremely High
Energy Cosmic Rays for a great workshop and hospitality. This work was
supported in part by the NSF through grant  AST-0071235 and DOE grant
DE-FG0291-ER40606 at the University of Chicago.

\end{document}